\documentclass[letter,bibyear]{aa} % for the letters

\usepackage{epsfig}
\usepackage{amsmath}
\usepackage{subfigure}
\usepackage{natbib}
\usepackage{multirow}
\usepackage{color}
\usepackage{longtable}
\usepackage{graphicx}
\usepackage{epstopdf}
\usepackage{booktabs}
\usepackage{txfonts}
\newcommand{\jms}{J.~Mol.~Spectrosc.}   % Journal of Molecular Spectroscopy
\newcommand{\jmst}{J.~Mol.~Struct.}   % Journal of Molecular Structure

\newcommand{\kms}{km s$^{-1}$}

\newcommand{\doce}{10$^{12}$\,cm$^{-2}$}
\newcommand{\trece}{10$^{13}$\,cm$^{-2}$}
\begin{document}

\title{Discovery of allenyl acetylene, H$_2$CCCHCCH, in TMC-1. A study of the isomers of C$_5$H$_4$
\thanks{Based on observations carried out with the Yebes 40m telescope
  (projects 19A003, 19A010, 20A014, 20D023).
  The 40m radiotelescope at Yebes Observatory is operated by the Spanish Geographic Institute
  (IGN, Ministerio de Transportes, Movilidad y Agenda Urbana).}}

\author{
J.~Cernicharo\inst{1},
C.~Cabezas\inst{1},
M.~Ag\'undez\inst{1},
B.~Tercero\inst{2,3},
N.~Marcelino\inst{1},
J.~R.~Pardo\inst{1},
F.~Tercero\inst{2},
J.D.~Gallego\inst{2},
J.A.~L\'opez-P\'erez\inst{2},
P.~deVicente\inst{2}
}

\institute{Grupo de Astrof\'isica Molecular, Instituto de F\'isica Fundamental (IFF-CSIC),
C/ Serrano 121, 28006 Madrid, Spain. \email jose.cernicharo@csic.es
\and Centro de Desarrollos Tecnol\'ogicos, Observatorio de Yebes (IGN), 19141 Yebes, Guadalajara, Spain.
\and Observatorio Astron\'omico Nacional (OAN, IGN), Madrid, Spain.
}

\date{Received; accepted}

\abstract{We present the discovery in TMC-1 of allenyl acetylene, H$_2$CCCHCCH, through
the observation of nineteen lines with a signal-to-noise ratio $\sim$4-15. For this species, 
we derived a rotational temperature of 7$\pm$1K and a column density of 1.2$\pm$0.2$\times$\trece.
The other well known isomer of this molecule, methyl diacetylene (CH$_3$C$_4$H), has also been
observed and we derived a similar rotational temperature, T$_r$=7.0$\pm$0.3 K, and
a column density for its two states ($A$ and $E$) of 6.5$\pm$0.3$\times$\doce. Hence,
allenyl acetylene and methyl diacetylene have a similar abundance. Remarkably, their 
abundances are close to that of vinyl acetylene (CH$_2$CHCCH).
We also searched for the other isomer of C$_5$H$_4$, HCCCH$_2$CCH (1.4-Pentadiyne), but
only a 3$\sigma$ upper limit of 2.5$\times$\doce\ to the column density can be established.
These results have been compared to state-of-the-art chemical models for TMC-1,
indicating the important role of these hydrocarbons in its chemistry.

The rotational parameters of allenyl acetylene have been improved by fitting the existing laboratory
data together with the frequencies of the transitions observed in TMC-1.
}


\titlerunning{CH2CCHCCH in TMC-1}
\authorrunning{Cernicharo et al.}

\maketitle

\section{Introduction}
More than 200 different chemical
species have been detected in space.
Most of them have a large dipole moment which permits an easy search
for their rotational transitions through radio astronomical observations.
However, only a few pure hydrocarbons, C$_n$H$_m$ (with m$\ge$2), have
been detected so far, and their
role in the chemistry of cold dark clouds is poorly understood. Molecules
such as C$_2$H$_2$, C$_2$H$_4$, and C$_2$H$_6$, which lack a permanent dipole moment, are studied through their derivatives, mainly
through the replacement of a hydrogen atom by the CN radical. In this context, it is worth noting that while
CH$_2$CHCN was detected in the early years of millimeter radio astronomy, its
equivalent with the CCH group, vinyl acetylene, has been detected only recently
towards the cold dark cloud TMC-1 with an abundance that is twice that of the
cyanide derivative \citep{Cernicharo2021a}. This is mainly due to the
huge difference in the dipole moments of CH$_2$CHCN ($\mu_a$=3.821\,D, \citealt{Krasnicki2011})
and CH$_2$CHCCH ($\mu_a$=0.43\,D, \citealt{Sobolev1962}). Hence, only a few hydrocarbons with a low
dipole moment have been found so far in the interstellar medium (ISM). Among them are propene \citep{Marcelino2007},
with a dipole moment of 0.36\,D \citep{Lide1957},
and deuterated methane, which has been tentatively detected towards the low mass protostar IRAS\,04368+2557 \citep{Sakai2012}. The later species 
has a very low dipole moment indeed, 0.0056\,D \citep{Ozier1969}. Pure unsaturated hydrocarbons radicals, C$_n$H,
and their anions, have moderate to very large dipole moments and have been detected up to $n = 8$ in interstellar and circumstellar clouds
\citep{Cernicharo1996,Remijan2007,Kawaguchi2007}.

In this letter, we report on the discovery of allenyl acetylene, H$_2$CCCHCCH (also named ethynyl allene), through
nineteen well detected rotational
transitions. This species was 
spectroscopically characterised \citep{McCarthy2020}, but never detected in space.
It is one of the possible isomers with molecular formula C$_5$H$_4$. We compare the derived abundance with that
of other acetylenic species such as CH$_3$C$_4$H (another C$_5$H$_4$ isomer), CH$_3$CCH, and CH$_2$CHCCH (the two
latter species are C$_4$H$_4$ isomers).
We also searched for another C$_5$H$_4$ isomer,
HCCCH$_2$CCH, but only upper limits are obtained.
These results are analysed in the context of a state-of-the-art chemical model of a cold dark cloud.

\section{Observations}
\label{observations}
New receivers, which were built within the Nanocosmos project\footnote{\texttt{https://nanocosmos.iff.csic.es/}}
and installed at the Yebes 40m radiotelescope, were used
for the observations of TMC-1. The Q-band receiver consists of two high electron mobility transistor cold
amplifiers, covering the
31.0-50.3 GHz range with horizontal and vertical polarisations. Receiver temperatures vary from 22 K at 32 GHz
to 42 K at 50 GHz. 
Eight 2.5 GHz wide fast Fourier transform spectrometers (FFTs), with a spectral resolution
of 38.15 kHz, provide the whole coverage of the Q-band in each polarization.
The main beam efficiency varies from 0.6 at
32 GHz to 0.43 at 50 GHz. A detailed description of the system is given by \citet{Tercero2021}.

The line survey of TMC-1 ($\alpha_{J2000}=4^{\rm h} 41^{\rm  m} 41.9^{\rm s}$ and $\delta_{J2000}=+25^\circ 41' 27.0''$)
in the Q-band was performed in several sessions. Previous results for the detection of C$_3$N$^-$ and C$_5$N$^-$
\citep{Cernicharo2020b}, HC$_5$NH$^+$ \citep{Marcelino2020}, HC$_4$NC \citep{Cernicharo2020c}, and HC$_3$O$^+$
\citep{Cernicharo2020a} were based on two observing runs performed in November 2019 and February 2020. Two
different frequency coverages were achieved, 31.08-49.52 GHz and 31.98-50.42 GHz, in order to check that no
spurious spectral ghosts were produced in the down-conversion chain,
which downconverts the signal from the receiver to 1-19.5 GHz and then splits it into 8 2.5 GHz bands which are
finally analysed by the FFTs. 
Additional data were taken in October and December 2020. A final observing run was performed in January 2021
to improve the line survey and to
further check the consistency of all observed spectral features. These new data have allowed the detection
of HC$_3$S$^+$ \citep{Cernicharo2021b} along with the acetyl cation, CH$_3$CO$^+$ \citep{Cernicharo2021c},
HDCCN \citep{Cabezas2021}, the isomers of C$_4$H$_3$N \citep{Marcelino2021}, and vinyl acetylene \citep{Cernicharo2021a}.

The observations were carried out using frequency switching with a frequency throw of 10\,MHz for the first two runs and of
8\,MHz for the later ones.
The intensity scale, antenna temperature ($T_A^*$), was calibrated using two absorbers
at different temperatures and the atmospheric transmission model (ATM; \citealt{Cernicharo1985};
\citealt{Pardo2001}). Calibration
uncertainties have been adopted to be 10~\%. The nominal spectral resolution of 38.15 kHz was used for the final
spectra. The sensitivity varies across the Q-band from 0.3 to 2.0 mK. All data have been analysed using the GILDAS
package\footnote{\texttt{http://www.iram.fr/IRAMFR/GILDAS}}.

\section{Results and discussion}
\label{results}
The sensitivity of our observations towards TMC-1 (see section \ref{observations}) is a factor of 10-20 better than
in previously published line surveys of this source at the same frequencies \citep{Kaifu2004}. This large
improvement has allowed us to detect a
forest of weak lines.
In fact, it has been possible to detect many individual lines \citep{Marcelino2021} from molecular species
that were reported previously only by stacking techniques \citep{Marcelino2021}.
Taking into account the large abundance found
in TMC-1 for cyanide derivatives of abundant species, and of the presence of nearly saturated hydrocarbons,
such as CH$_3$CHCH$_2$
\citep{Marcelino2007} and CH$_2$CHCCH \citep{Cernicharo2021a},
we searched for similar species containing CCH such as H$_2$CCCHCCH and HCCCH$_2$CCH. We compared the derived abundances with that of the well know species in this source 
methyl diacetylene, CH$_3$C$_4$H \citep{MacLeod1984,Walmsley1984}.
Line identifications in this TMC-1 survey were performed using the MADEX catalogue \citep{Cernicharo2012},
the Cologne Database of Molecular Spectroscopy catalogue (CDMS; \citealt{Muller2005}), and the
JPL catalogue \citep{Pickett1998}.

\begin{figure*}[]
\centering
\includegraphics[scale=0.9,angle=0]{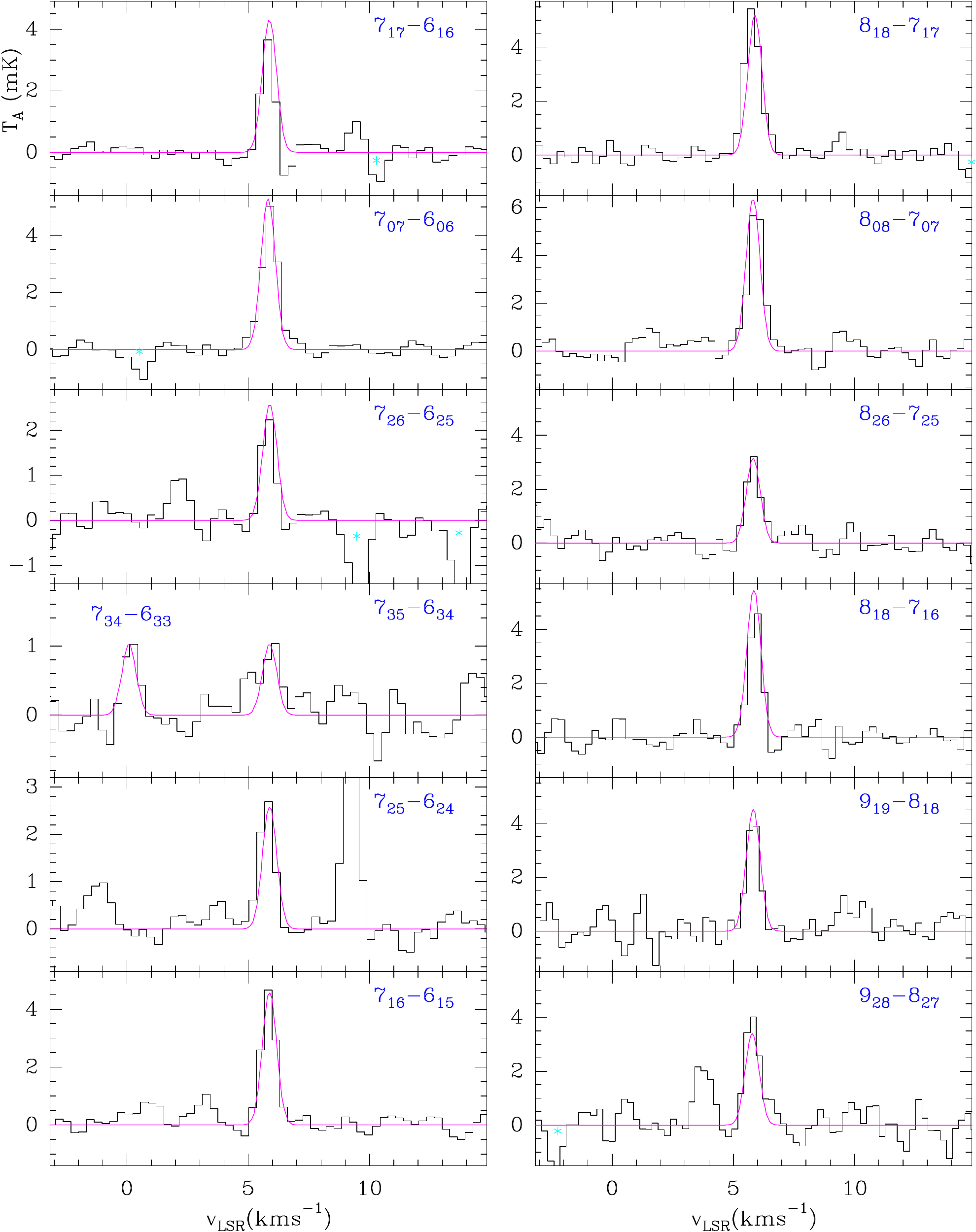}
\caption{Observed transitions of H$_2$CCCHCCH towards TMC-1.
The abscissa corresponds to the rest frequency of the lines assuming a
local standard of rest velocity of the source of 5.83 km s$^{-1}$. Frequencies and intensities for the observed lines
are given in Table \ref{obs_line_parameters}.
The ordinate is the antenna temperature, corrected for atmospheric and telescope losses, in 
millikelvin.
The violet line shows the computed synthetic spectrum for this species for T$_r$=7 K and
a column density of 1.2$\times$\trece. Cyan stars indicate the position of negative features produced
in the folding of the frequency switching observations.
}
\label{fig_ch2cchcch}
\end{figure*}

\subsection{Allenyl acetylene, H$_2$CCCHCCH}
\label{CH2CCHCCH}
Allenyl acetylene is one of the possible C$_5$H$_4$ isomers. We
calculated the relative energies of its three most stable isomers, namely H$_2$CCCHCCH, CH$_3$C$_4$H, and HCCCH$_2$CCH.
Structural optimisation calculations for the lowest energy conformers of each isomer
were done using the M{\o}ller-Plesset post-Hartree-Fock method \citep{Moller1934} and explicitly
electron correlation effects through perturbation theory up to the second and
the Dunning's consistent polarised valence triple-$\zeta$ basis set (MP2/cc-pVTZ)
\citep{Dunning1989}. These calculations were performed using the Gaussian 09
programme package \citep{Frisch2009}. Our results show that CH$_3$C$_4$H is the global minimum, while H$_2$CCCHCCH
and HCCCH$_2$CCH lie at 1.44~kJ\,mol$^{-1}$ and 1.62~kJ\,mol$^{-1}$, respectively, over CH$_3$C$_4$H.

Spectroscopic constants for H$_2$CCCHCCH were derived from a fit to the lines reported by \cite{McCarthy2020} and
implemented in the MADEX code \citep{Cernicharo2012}. Nineteen lines with $K_a$=0, 1, 2, and 3 have been detected in TMC-1. A
selected sample of them is
shown in Fig. \ref{fig_ch2cchcch}. Derived frequencies and line parameters are given in Table \ref{obs_line_parameters}.
All lines of allenyl acetylene that are not blended with lines from other species and with expected intensities above 1\,mK
were detected in our survey.
Four additional features, with expected intensities of 1-3 mK, fall in the middle of a forest of lines produced by
H$_2$CCN \citep{Cabezas2021} so that deriving their frequencies and intensities was unreliable.

In order to compute column densities, we calculated the electric dipole moment components of H$_2$CCCHCCH at the MP2/cc-pVTZ
level of theory. 
The $|\mu_{a}|$ and $|\mu_{a}|$ derived values are 0.630 D and 0.011 D, respectively.
They are in agreement with those previously reported by \cite{Lee2019},
which were obtained using a density functional theory level of theory (M05/6-31G(d)).

An analysis of the observed intensities through a rotational diagram
provides a rotational temperature of 9$\pm$1\,K. We performed a model fitting directly
to the observed line profiles as described  by \citet{Cernicharo2021a}, with the result that
the best match between the computed synthetic spectrum and the observations was obtained for
T$_r$=7\,K and N(H$_2$CCCHCCH)=(1.2$\pm$0.2)$\times$\trece. Figure \ref{fig_ch2cchcch} shows
the synthetic spectrum. Only the transition {$9_{19}-8_{1,8}$ required a correction
intensity by a factor of 0.8, while for all the other lines the model matches the
observations perfectly.
Using the H$_2$ column density derived by \citet{Cernicharo1987}, the abundance of H$_2$CCCHCCH relative
to H$_2$ towards TMC-1 is 1.2$\times$10$^{-9}$. This abundance is
similar to that of vinyl acetylene \citep{Cernicharo2021a}, about $\sim$3 below that
of propylene (CH$_3$CHCH$_2$; \citealt{Marcelino2007}), and a factor of ten below that of methyl
acetylene (CH$_3$CCH; \citealt{Cabezas2021}).
Hence, allenyl acetylene is one of the most
abundant hydrocarbons in TMC-1, and probably, together with CH$_3$C$_4$H (see Sect. \ref{CH3C4H}),
the most abundant compound containing five carbon atoms.
It is interesting
to compare the abundance of allenyl acetylene with that of cyano allene (H$_2$CCCHCN).
This species has been recently analysed by \citet{Marcelino2021}, resulting in a rotational temperature
of 5.5$\pm$0.3 K and a column density of (2.7$\pm$0.2)$\times$\doce.
Consequently, the abundance ratio of H$_2$CCCHCCH over H$_2$CCCHCN is $\sim$4.5,
that is to say the acetylenic derivative
of allene is slightly more abundant than the cyanide one. A similar value ($\sim$1.8) was obtained for
the abundance ratio between CH$_2$CHCCH and CH$_2$CHCN \citep{Cernicharo2021a}.

The measured frequencies of the lines observed in TMC-1 can be used to improve the rotational and distortion
constants of H$_2$CCCHCCH.
We used the fitting code FITWAT described in \citet{Cernicharo2018}. Table \ref{rotational_constants}
provides the results obtained by fitting the laboratory data of \citet{McCarthy2020} alone, and those
obtained from a fit to the merged laboratory plus the TMC-1 frequencies. A significant improvement in the uncertainty in
the rotational and distortion constants was obtained. The fit to the laboratory data alone results in exactly
the same constants as those obtained by \citet{McCarthy2020}. The merged fit is recommended to predict the
frequency of the rotational transitions of this species above 50 GHz. Predictions could be accurate enough
up to 150 GHz to allow for a search of this species in the 3-mm domain. Table \ref{obs_frequencies} provides the observed and calculated
frequencies and their differences.

\begin{table}
\tiny
\caption{Rotational and distortion constants of H$_2$CCCHCCH.}
\label{rotational_constants}
\centering
\begin{tabular}{lcc}
\hline
Constant            & Laboratory$^a$      & This work         \\
$A$ (MHz)           &  25963.54(166)      &25961.178(785)     \\
$B$ (MHz)           &   2616.375797(314)  & 2616.376200(221)  \\
$C$ (MHz)           &   2412.573364(286)  & 2412.573306(194)  \\
$\Delta_J$ (kHz)    &      1.15462(391)   &   1.15734(112)    \\
$\Delta_{JK}$ (kHz) &    -85.5217(356)    & -85.4904(279)     \\
$\delta_J$ (kHz)    &      0.28161(457)   &   0.28598(132)    \\
\hline
Number of lines     & 14                  & 33                \\
$\sigma$(kHz)       & 1.0                 & 8.5               \\
$J_{max}$, $Ka_{max}$& 5, 2                & 10,3              \\
$\nu_{max}$ (GHz)   & 25.144              & 49.218            \\
\hline
\hline
\end{tabular}
\tablefoot{\\
\tablefoottext{a}{Laboratory frequencies from \citet{McCarthy2020}.}\\
\\
}
\end{table}

\subsection{Methyl diacetylene, CH$_3$C$_4$H}
\label{CH3C4H}
This C$_5$H$_4$ isomer was found in TMC-1 by \cite{MacLeod1984} and \cite{Walmsley1984}. We used spectroscopic 
information from \cite{Bester1984}, \cite{Heath1955}, and \cite{Cazzoli2008} to
obtain the rotational constants that were implemented in MADEX \citep{Cernicharo2012}.
The dipole moment, 1.207 D, was measured by \citet{Bester1984}.
The constants $A$ and $D_{\rm K}$ are taken from \citet{Muller2002}.
Five rotational transitions, from $J_u=8$ up $J_u=12$, lie within our line survey. The $K$ = 0, 1, and 2 components 
of these
transitions were observed as shown in Fig. \ref{fig_ch3c4h}.
Derived line parameters are given in Table \ref{obs_line_ch3c4h}.
The detection of the $K$=2 component
allows for a good determination of the rotational temperature.
As for H$_2$CCCHCCH, we assumed a uniform brightness source with a radius of 40$''$ \citep{Fosse2001}.
In a first step, we derived T$_r$ from a rotational diagram of the lines of the $A$ and $E$ species. A
similar rotation temperature of 7$\pm$1\,K was obtained for both symmetry species. Then, we produced
a synthetic spectrum that was compared with the observed line profiles, allowing us to refine the derived parameters.
We found that a synthetic spectrum with T$_r$=7\,K and N($A$-CH$_3$C$_4$H)=N($E$-CH$_3$C$_4$H)=(6.5$\pm$0.2)$\times$\doce\,
matches the observed spectra very well (see Fig. \ref{fig_ch3c4h}). The synthetic spectrum was corrected for beam
dilution and beam efficiency. Consequently, the total column density of methyl diacetylene is (1.3$\pm$0.4)$\times$\trece.
This value is in good agreement with those derived by \cite{Walmsley1984} and \cite{MacLeod1984}
when corrected for the different dipole moment used by these authors (1.0, and 0.9 D, respectively).
Hence, both species, allenyl acetylene and methyl diacetylene, have similar abundances in TMC-1. They are
only three times less abundant than propene \citep{Marcelino2007} and ten times less abundant than methyl acetylene \citep{Cabezas2021}.

\begin{figure}[]
\centering
\includegraphics[scale=0.65,angle=0]{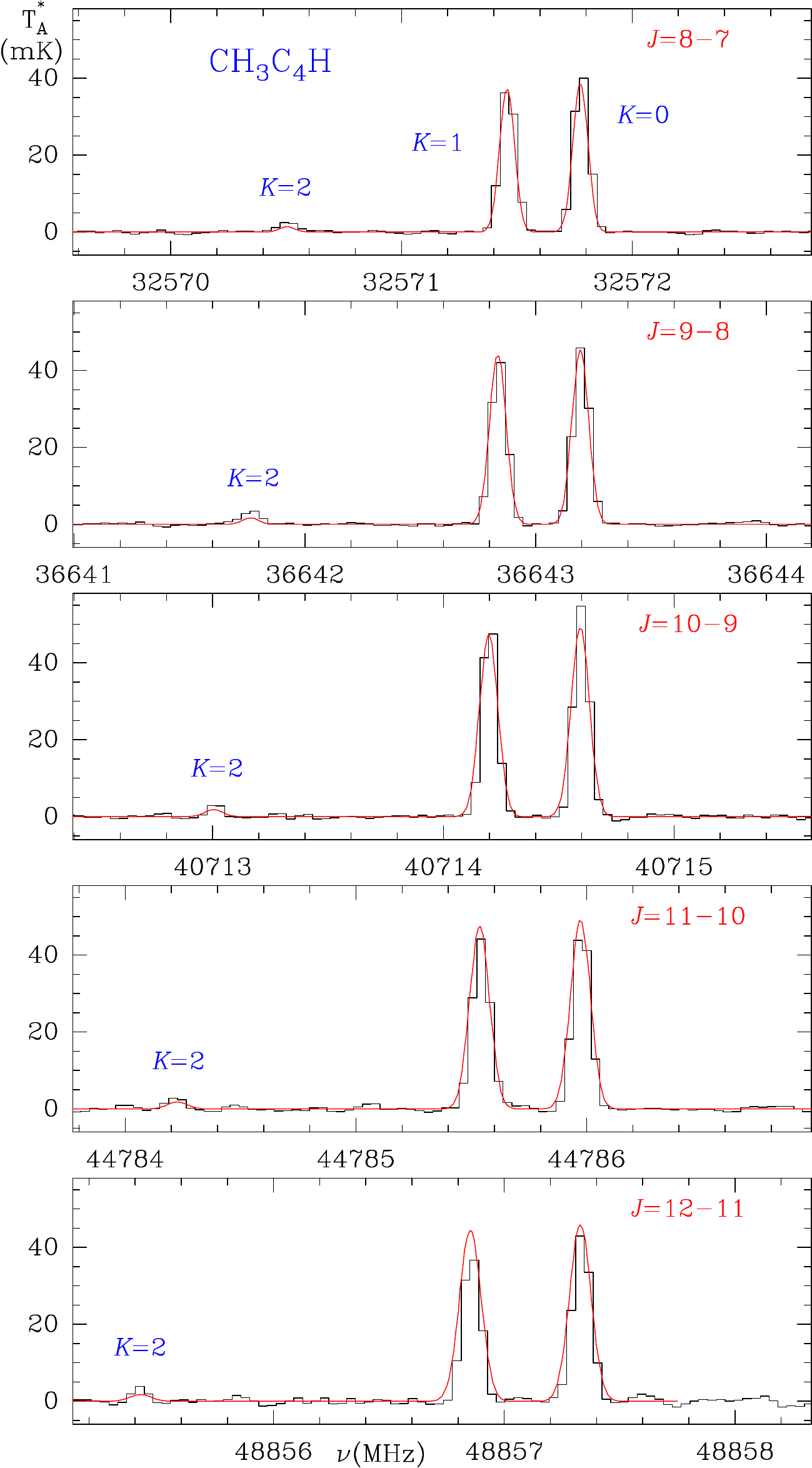}
\caption{Observed transitions of CH$_3$C$_4$H towards TMC-1.
The abscissa corresponds to the rest frequency of the lines assuming a
local standard of rest velocity of TMC-1 of 5.83 km s$^{-1}$. Line parameters for the observed transitions
are given in Table \ref{obs_line_parameters}.
The ordinate is the antenna temperature, corrected for atmospheric and telescope losses, in millikelvin.
The red line shows the computed synthetic spectrum for this species for T$_r$=7 K and
a column density of 6.5$\times$\doce\, for each of the $A$ and $E$ states of
vinyl diacetylene.
}
\label{fig_ch3c4h}
\end{figure}

\subsection{HCCCH$_2$CCH}
\label{HCCCH2CCH}
We note that 1,4-Pentadiyne, HCCCH$_2$CCH, is another C$_5$H$_4$ isomer. Its rotational spectrum was measured
in the laboratory by \citet{Kuczkowski1981}.
Its dipole moment, 0.52 D, was measured by the same authors. We searched
for it through more than ten rotational transitions falling in the 31-50 GHz range.
None of them were detected. We derived a 3$\sigma$ upper limit to its column density of
4$\times$\doce. This moderate upper limit is due to the relatively low dipole moment of this molecule.

\subsection{Discussion} \label{discussion}

\begin{figure}
\centering
\includegraphics[angle=0,width=\columnwidth]{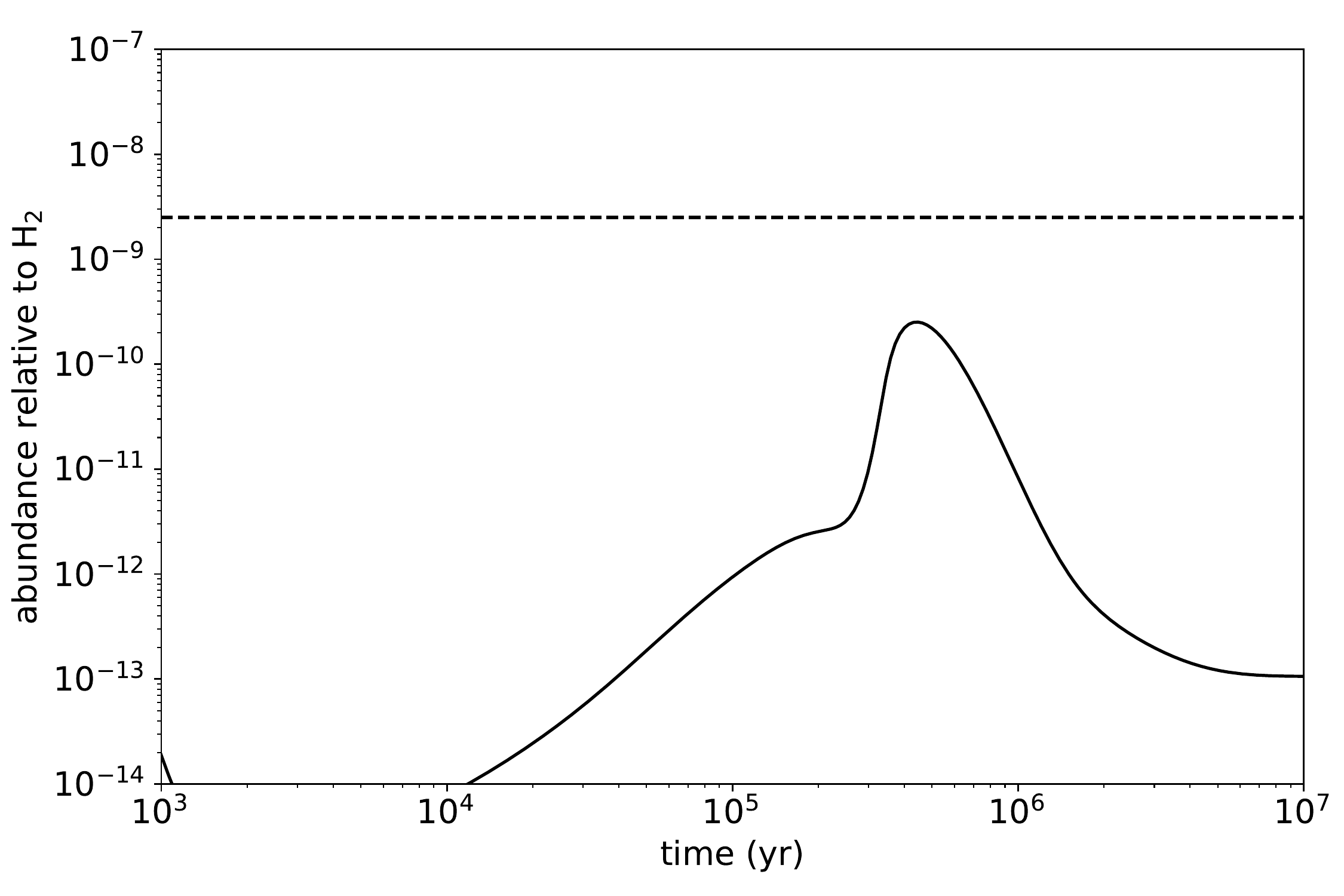}
\caption{Calculated fractional abundance of C$_5$H$_4$ (allowing for various isomers) as a function of time. 
The horizontal dashed line corresponds to the sum of the observed abundances of the two C$_5$H$_4$ 
isomers (CH$_3$C$_4$H and H$_2$CCCHCCH) detected in TMC-1.}
\label{fig:chem_model}
\end{figure}

As a matter of fact, the two most stable C$_5$H$_4$ isomers were detected in TMC-1 and both have similar 
abundances. To learn about the formation of these molecules under cold dark cloud conditions, we carried 
out chemical model calculations similar to those presented in \cite{Marcelino2021}. For chemical model 
purposes, we considered that the species with molecular formula C$_5$H$_4$ accounts for the various possible isomers. 
According to our chemical model, the peak abundance of C$_5$H$_4$ isomers under cold dark cloud conditions is 
$2.5 \times 10^{-10}$ relative to H$_2$, which is ten times smaller than the sum of the observed abundances 
of CH$_3$C$_4$H and H$_2$CCCHCCH in TMC-1 (see Fig.~\ref{fig:chem_model}). While the difference between the calculated 
and observed abundance is significant, it is interesting to inspect which are the main formation routes in the 
chemical model. Reactions of the CCH radical with methyl acetylene (CH$_3$CCH) and allene (H$_2$CCCH$_2$)
\begin{subequations} \label{reac:cch+ch3cch}
\begin{align}
\rm CCH + CH_3CCH & \rightarrow \rm CH_3C_4H + H, \label{reac:cch+ch3cch_a} \\
                                    & \rightarrow \rm H_2CCCHCCH + H, \label{reac:cch+ch3cch_b}
\end{align}
\end{subequations}
\begin{subequations} \label{reac:cch+ch2cch2}
\begin{align}
\rm CCH + CH_2CCH_2 & \rightarrow \rm CH_3C_4H + H, \label{reac:cch+ch2cch2_a} \\
                                        & \rightarrow \rm H_2CCCHCCH + H, \label{reac:cch+ch2cch2_b}
\end{align}
\end{subequations}
account for most of the C$_5$H$_4$ isomers formation. These reactions were experimentally found to be rapid at low temperatures, down to 63~K \citep{Carty2001}, 
and they are probably also fast at temperatures around 10 K. The reaction of C$_2$ with propylene, 
which has also been measured to be rapid down to 77~K \citep{Daugey2008}, is also an important 
source of C$_5$H$_4$ isomers, while a third route involving the dissociative recombination of 
the ion C$_5$H$_5^+$ does also contribute to their formation.

Although the branching ratios of reactions (\ref{reac:cch+ch3cch}) and (\ref{reac:cch+ch2cch2}) are not 
precisely known, methyl acetylene and allenyl acetylene are the most likely products \citep{Kaiser2001,Zhang2009,Goulay2011}. 
In fact, it would be interesting to verify if CH$_3$C$_4$H is the preferred product of reaction (\ref{reac:cch+ch3cch}) 
and if H$_2$CCCHCCH is preferentially formed in reaction (\ref{reac:cch+ch2cch2}); this would allow one to probe the abundance 
of the non-polar hydrocarbon allene, which is expected to be large. The chemical scheme depicted by reactions 
(\ref{reac:cch+ch3cch}) and (\ref{reac:cch+ch2cch2}) 
is similar to that driving the formation of C$_4$H$_3$N isomers 
in which reactions of the CN radical with CH$_3$CCH and H$_2$CCCH$_2$ are at the heart of the synthesis of the various 
C$_4$H$_3$N isomers, as discussed by \cite{Marcelino2021}. In fact, it is worth noting that the abundance ratio 
C$_5$H$_4$/C$_4$H$_3$N of 3.5 (\citealt{Marcelino2021} and this study) observed in TMC-1 is not far from the CCH/CN abundance 
ratio of 10 observed in this source \citep{Pratap1997}.

\begin{acknowledgements}

We thank Ministerio de Ciencia e Innovaci\'on of Spain (MICIU) for funding support through projects
AYA2016-75066-C2-1-P, PID2019-106110GB-I00, PID2019-107115GB-C21 / AEI / 10.13039/501100011033, and
PID2019-106235GB-I00. We also thank ERC for funding
through grant ERC-2013-Syg-610256-NANOCOSMOS. M.A. thanks MICIU for grant RyC-2014-16277.
\end{acknowledgements}

\normalsize

\begin{appendix}
\section{Line parameters of H$_2$CCHCCH and CH$_3$C$_4$H}
Line parameters for the different molecules studied in this work were obtained by fitting a Gaussian line
profile to the observed data. A window of $\pm$ 20 \kms\, around the v$_{LSR}$ of the source was
considered for each transition. The derived line parameters for H$_2$CCCHCCH are given in Table \ref{obs_line_parameters}.
Those of CH$_3$C$_4$H are provided in Table \ref{obs_line_ch3c4h}.

\begin{table}
\tiny
\caption{Observed line parameters for CH2CCHCCH in TMC-1.}
\label{obs_line_parameters}
\centering
\begin{tabular}{ccccc}
\hline
$J K_a K_c$        &$\nu_{obs}$~$^a$    & $\int T_A^* dv$~$^b$ & $\Delta v$~$^c$ & $T_A^*$ \\
                   &  (MHz)             & (mK\,km\,s$^{-1}$)   & (km\,s$^{-1}$)  & (mK)    \\
\hline
$ 7_{1, 7}-6_{1,6}$& 34472.262(10)& 2.4$\pm$0.4        & 0.60$\pm$0.05& 3.7$\pm$0.3$^d$ \\ %7.64E-03
$ 7_{0, 7}-6_{0,6}$& 35126.923(10)& 4.0$\pm$0.5        & 0.77$\pm$0.05& 4.8$\pm$0.3     \\ %8.78E-03
$ 7_{2, 6}-6_{2,5}$& 35195.532(10)& 1.6$\pm$0.4        & 0.63$\pm$0.07& 2.4$\pm$0.3     \\ %5.15E-03                                                   &  \\ %5.15E-03
$ 7_{3, 5}-6_{3,4}$& 35222.531(10)& 1.0$\pm$0.4        & 0.70$\pm$0.20& 1.3$\pm$0.3     \\ %2.61E-3                                                    &  \\ %2.61E-03
$ 7_{3, 4}-6_{3,3}$& 35223.195(10)& 0.7$\pm$0.3        & 0.60$\pm$0.17& 1.0$\pm$0.3     \\ %2.61E-03
$ 7_{2, 5}-6_{2,4}$& 35269.664(10)& 2.2$\pm$0.5        & 0.74$\pm$0.08& 2.7$\pm$0.3     \\ %5.16E-03
$ 7_{1, 6}-6_{1,5}$& 35897.418(10)& 3.6$\pm$0.5        & 0.73$\pm$0.05& 4.7$\pm$0.3     \\ %7.76E-03
$ 8_{1, 8}-7_{1,7}$& 39390.116(10)& 4.3$\pm$0.4        & 0.77$\pm$0.04& 5.3$\pm$0.3     \\ %8.37E-03
$ 8_{0, 8}-7_{0,7}$& 40118.252(10)& 4.7$\pm$0.4        & 0.72$\pm$0.05& 6.1$\pm$0.3     \\ %9.55E-03
$ 8_{2, 7}-7_{2,6}$& 40218.511(20)& 0.8$\pm$0.4        & 0.50$\pm$0.20& 1.6$\pm$0.3$^e$ \\ %5.72E-03
$ 8_{2, 6}-7_{2,5}$& 40329.448(10)& 2.6$\pm$0.6        & 0.75$\pm$0.09& 3.2$\pm$0.4     \\ %5.73E-03
$ 8_{1, 7}-7_{1,6}$& 41018.002(10)& 2.7$\pm$0.5        & 0.55$\pm$0.06& 4.6$\pm$0.4     \\ %8.45E-03
$ 9_{1, 9}-8_{1,8}$& 44305.369(10)& 2.8$\pm$0.5        & 0.61$\pm$0.07& 4.3$\pm$0.5     \\ %8.68E-03
$ 9_{0, 9}-8_{0,8}$& 45099.094(20)& 4.4$\pm$1.0        & 0.57$\pm$0.02& 8.0$\pm$2.0$^f$ \\ %9.82E-03
$ 9_{2, 8}-8_{2,7}$& 45239.452(10)& 3.2$\pm$0.8        & 0.73$\pm$0.10& 4.1$\pm$0.6     \\ %5.96E-03
$ 9_{3, 7}-8_{3,6}$& 45291.918(10)&                    &              &    $g$          \\ %3.18E-03
$ 9_{3, 6}-8_{3,5}$& 45294.382(20)& 1.3$\pm$0.5        & 0.69$\pm$0.17& 1.9$\pm$0.5     \\ %3.18E-03
$ 9_{2, 7}-8_{2,6}$& 45397.522(20)& 1.5$\pm$0.5        & 0.52$\pm$0.15& 1.8$\pm$0.6$^h$ \\ %5.97E-03
$ 9_{1, 8}-8_{1,7}$& 46135.431(10)& 2.3$\pm$0.5        & 0.60$\pm$0.07& 4.7$\pm$0.6     \\ %8.67E-03
$10_{1,10}-9_{1,9}$& 49217.784(10)& 3.2$\pm$0.7        & 0.65$\pm$0.10& 4.9$\pm$0.6     \\ %8.55E-03
\hline
\end{tabular}
\tablefoot{\\
\tablefoottext{a}{Observed frequency assuming a v$_{LSR}$ of 5.83 \kms.}\\
\tablefoottext{b}{Integrated line intensity in mK\,km\,s$^{-1}$.}\\
\tablefoottext{c}{Linewidth at half intensity derived by fitting a Gaussian function to
the observed line profile (in km\,s$^{-1}$).}\\
\tablefoottext{d}{Blended with a negative feature produced in the folding of the frequency switching observing procedure.}\\
\tablefoottext{e}{Blended with two negative features produced in the frequency switching folding. The intensity is uncertain}\\
\tablefoottext{f}{Heavily blended with HCCCH$_2$CN. Frequency difference between both lines lower than 20 kHz. We estimate
that the contribution of this contaminating feature, by comparison with other similar lines of this
species, is $\sim$50\%}\\
\tablefoottext{g}{Fully blended with a negative feature produced in the frequency switching folding. The fit is unreliable.}\\
\tablefoottext{h}{Blended with two lines from other species. The separation between them still allows for a reasonable
fit to the line parameters.}\\
}
\end{table}

\begin{table}
\tiny
\caption{Observed line parameters for CH$_3$C$_4$H in TMC-1.}
\label{obs_line_ch3c4h}
\centering
\begin{tabular}{{ccccccc}}
\hline
 $J_{\rm u}$ & $K$ & $\nu$ &$\int T_{\rm A}^* dv$$^a$ & $V_{\rm LSR}$$^b$ & $\Delta$v$^c$ & $T_{\rm A}^*$$^d$\\
             &     & (MHz)      &(mK km s$^{-1}$)       & (km s$^{-1}$) & (km s$^{-1}$)  & (mK)      \\
\hline
 8 &  2      &  32570.504 & $ 2.1\pm0.5$  & $5.80\pm0.05$ & $0.82\pm0.11$  & $ 2.4\pm0.3$\\
 8 &  1      &  32571.457 & $30.8\pm0.7$  & $5.80\pm0.02$ & $0.74\pm0.01$  & $39.3\pm0.3$\\
 8 &  0      &  32571.775 & $33.6\pm0.7$  & $5.80\pm0.02$ & $0.73\pm0.01$  & $42.8\pm0.3$\\
 9 &  2      &  36641.764 & $ 2.8\pm0.5$  & $5.81\pm0.03$ & $0.79\pm0.07$  & $ 3.4\pm0.3$\\
 9 &  1      &  36642.836 & $32.0\pm0.7$  & $5.79\pm0.04$ & $0.69\pm0.01$  & $43.8\pm0.3$\\
 9 &  0      &  36643.194 & $33.9\pm0.7$  & $5.80\pm0.02$ & $0.69\pm0.01$  & $46.3\pm0.3$\\
10 &  2      &  40713.005 & $ 2.0\pm0.5$  & $5.78\pm0.03$ & $0.47\pm0.08$  & $ 3.9\pm0.4$\\
10 &  1      &  40714.197 & $31.9\pm0.7$  & $5.81\pm0.02$ & $0.57\pm0.01$  & $53.0\pm0.4$\\
10 &  0      &  40714.594 & $34.2\pm0.7$  & $5.80\pm0.02$ & $0.58\pm0.01$  & $55.2\pm0.4$\\
11 &  2      &  44784.226 & $ 2.2\pm0.7$  & $5.89\pm0.05$ & $0.61\pm0.10$  & $ 3.4\pm0.5$\\
11 &  1      &  44785.536 & $30.0\pm0.8$  & $5.79\pm0.02$ & $0.63\pm0.01$  & $44.5\pm0.5$\\
11 &  0      &  44785.973 & $30.9\pm0.8$  & $5.80\pm0.02$ & $0.60\pm0.02$  & $48.6\pm0.5$\\
12 &  2      &  48855.424 & $ 1.9\pm0.6$  & $5.85\pm0.04$ & $0.44\pm0.07$  & $ 4.1\pm0.6$\\
12 &  1      &  48856.853 & $23.8\pm0.8$  & $5.80\pm0.02$ & $0.57\pm0.01$  & $39.1\pm0.6$\\
12 &  0      &  48857.330 & $27.1\pm0.8$  & $5.80\pm0.02$ & $0.58\pm0.01$  & $44.2\pm0.6$\\
\hline
\end{tabular}
\tablefoot{
\tablefoottext{a}{Observed frequency assuming a v$_{LSR}$ of 5.83 \kms.}\\ 
\tablefoottext{b}{Local standard of rest velocity of the line in km\,s$^{-1}$.}\\ 
\tablefoottext{c}{Integrated line intensity in mK\,km\,s$^{-1}$.}\\ 
\tablefoottext{d}{Linewidth at half intensity derived by fitting a Gaussian function to 
the observed line profile (in km\,s$^{-1}$).}\\ 
     }
\end{table}
\normalsize

\section{Observed and calculated frequencies for H$_2$CCHCCH}
In this section, we provide the full list of observed and calculated
frequencies for allyl cyanide. The laboratory data come from \citet{McCarthy2020}.
All lines were weighted in the fit as 1/$\sigma^2$. Laboratory data have
an accuracy of 2 kHz, while the corresponding uncertainty for the rotational
transitions measured in TMC-1 is 10 kHz, with the exception of four lines showing
some partial blending with other features for which we assigned an uncertainty of 20 kHz.

\begin{table*}
\small
\caption{Observed frequencies for H$_2$CCCHCCH in the laboratory and in TMC-1.}
\label{obs_frequencies}
\centering
\begin{tabular}{ccccccc}
\hline
  Transition       &    $\nu$  & Unc.&$\nu_{cal}$& Unc. & $\nu(obs-cal)$& \\
                   &  (MHz)    &(MHz)& (MHz)     &(MHz) &   (MHz)       & \\
\hline
  $ 2_{1,2}-1_{1,1}$&  9854.409&0.002& 9854.4102 &0.0004&   -0.0012     & A\\
  $ 2_{0,2}-1_{0,1}$& 10056.532&0.002&10056.5335 &0.0003&   -0.0015     & A\\
  $ 2_{1,1}-1_{1,0}$& 10261.996&0.002&10261.9977 &0.0005&   -0.0017     & A\\
  $ 3_{1,3}-2_{1,2}$& 14780.736&0.002&14780.7355 &0.0006&    0.0005     & A\\
  $ 3_{0,3}-2_{0,2}$& 15081.411&0.002&15081.4106 &0.0004&    0.0004     & A\\
  $ 3_{2,2}-2_{2,1}$& 15088.775&0.002&15088.7753 &0.0005&   -0.0003     & A\\
  $ 3_{1,2}-2_{1,1}$& 15392.077&0.002&15392.0770 &0.0007&    0.0000     & A\\
  $ 4_{1,4}-3_{1,3}$& 19706.013&0.002&19706.0129 &0.0007&    0.0001     & A\\
  $ 4_{0,4}-3_{0,3}$& 20102.224&0.002&20102.2239 &0.0006&    0.0001     & A\\
  $ 4_{2,3}-3_{2,2}$& 20117.206&0.002&20117.2044 &0.0006&    0.0016     & A\\
  $ 4_{2,2}-3_{2,1}$& 20130.481&0.002&20130.4822 &0.0006&   -0.0012     & A\\
  $ 4_{1,3}-3_{1,2}$& 20521.046&0.002&20521.0456 &0.0008&    0.0004     & A\\
  $ 5_{1,5}-4_{1,4}$& 24629.906&0.002&24629.9066 &0.0007&   -0.0006     & A\\
  $ 5_{2,4}-4_{2,3}$& 25144.638&0.002&25144.6375 &0.0007&    0.0005     & A\\
  $ 7_{1,7}-6_{1,6}$& 34472.262&0.010&34472.2569 &0.0013&    0.0051     & B\\
  $ 7_{0,7}-6_{0,6}$& 35126.923&0.010&35126.9230 &0.0020&    0.0000     & B\\
  $ 7_{2,6}-6_{2,5}$& 35195.532&0.010&35195.5240 &0.0011&    0.0080     & B\\
  $ 7_{3,5}-6_{3,4}$& 35222.531&0.010&35222.5236 &0.0025&    0.0074     & B\\
  $ 7_{3,4}-6_{3,3}$& 35223.195&0.010&35223.2044 &0.0024&   -0.0094     & B\\
  $ 7_{2,5}-6_{2,4}$& 35269.664&0.010&35269.6575 &0.0024&    0.0065     & B\\
  $ 7_{1,6}-6_{1,5}$& 35897.418&0.010&35897.4116 &0.0019&    0.0064     & B\\
  $ 8_{1,8}-7_{1,7}$& 39390.116&0.010&39390.1081 &0.0021&    0.0079     & B\\
  $ 8_{0,8}-7_{0,7}$& 40118.252&0.010&40118.2535 &0.0030&   -0.0015     & B\\
  $ 8_{2,7}-7_{2,6}$& 40218.511&0.020&40218.4833 &0.0016&    0.0277     & B\\
  $ 8_{2,6}-7_{2,5}$& 40329.448&0.010&40329.4495 &0.0038&   -0.0015     & B\\
  $ 8_{1,7}-7_{1,6}$& 41018.002&0.010&41017.9994 &0.0029&    0.0026     & B\\
  $ 9_{1,9}-8_{1,8}$& 44305.369&0.010&44305.3697 &0.0031&   -0.0007     & B\\
  $ 9_{0,9}-8_{0,8}$& 45099.094&0.020&45099.0983 &0.0044&   -0.0043     & B\\
  $ 9_{2,8}-8_{2,7}$& 45239.452&0.010&45239.4588 &0.0023&   -0.0068     & B\\
  $ 9_{3,6}-8_{3,5}$& 45294.382&0.020&45294.4077 &0.0034&   -0.0257     & B\\
  $ 9_{2,7}-8_{2,6}$& 45397.522&0.020&45397.5168 &0.0055&    0.0052     & B\\
  $ 9_{1,8}-8_{1,7}$& 46135.431&0.010&46135.4358 &0.0043&   -0.0048     & B\\
  $10_{1,10}-9_{1,9}$&49217.784&0.010&49217.7869 &0.0045&   -0.0029     & B\\
\hline
\end{tabular}
\tablefoot{\\
\tablefoottext{A}{Observed frequencies in the laboratory are from \citet{McCarthy2020}.} \\
\tablefoottext{B}{Frequencies observed in TMC-1 adopting a v$_{\rm LSR}$ of 5.83 km s$^{-1}$.}\\
\tablefoottext{b}{Observed minus calculated frequencies in MHz.}\\
}
\end{table*}

\end{appendix}

\end{document}